\documentclass[11pt]{article}

\usepackage{amsmath}
\usepackage{graphicx}
\usepackage{indentfirst}
\usepackage{amssymb}
\usepackage{cite}
\usepackage{color}
\usepackage{subfigure}

\begin{document}

\title{ Thermodynamics and weak cosmic censorship conjecture of charged AdS black hole in the Rastall gravity with pressure}

\date{}
\maketitle

\begin{center}
\author{Xin-Yun Hu,}$^{\textbf{1},}$\footnote{huxinyun@126.com}
\author{Ke-Jian He,}$^{\textbf{2},}$\footnote{kjhe94@163.com }
\author{Zhong-Hua Li,}$^{\textbf{2},}$\footnote{sclzh888@163.com}
\author{Guo-Ping Li}$^{*\textbf{2},}$\footnote{Corresponding author: gplipys@yeah.net}

\vskip 0.25in
$^{\textbf{1}}$\it{College of Economic and Management, Chongqing Jiaotong University, Chongqing 400074, China}\\
$^{\textbf{2}}$\it{Physics and Space College, China West Normal University, Nanchong 637000, China}

\end{center}
\vskip 0.6in
{\abstract
{
   Treating the cosmological constant as a dynamical variable,  we  investigate the thermodynamics and weak cosmic censorship conjecture (WCCC) of a charged AdS black hole (BH) in the Rastall gravity. We determine the  energy  momentum relation of  charged fermion at the horizon of the   BH  by using  the Dirac equation. Based on this relation,  we show that the first law of thermodynamics (FLT) still holds as a  fermion is absorbed by the  BH. However, the entropy of both the extremal and near-extremal  BH  decreases in the irreversible process, which means that  the second law of thermodynamics (SLT) is violated.  Furthermore, we verify the validity of the   WCCC   by the minimum values of the metric function $h(r)$ at its final state.  For the extremal charged AdS  BH in the Rastall gravity, we find that the   WCCC   is valid always since the   BH is extreme. While for the case of near-extremal  BH,  we find  the   WCCC   could be violable in the extended phase space (EPS), depending on the value of the parameters of the   BH and their variations. }
}

\thispagestyle{empty}
\newpage
\setcounter{page}{1}

\section{Introduction}\label{sec1}
An important  feature for a   BH  is that it owns an  event horizon, which is a boundary where any  particle, even light,  cannot escape from it. The event horizon also plays a critical role in hiding the inner structure of  BHs, including the curvature singularity. Near the singularity, the curvature of spacetime tends to diverge and the laws of physics are broken. To avoid these phenomena,  Penrose proposed the   WCCC,  according to which,  the singularities resulting from gravitational collapse are hidden by the  BH event horizon for  the observers located at infinity\cite{ref1,ref2}. However, there is no universal proof of this conjecture until now, so we should check it one by one in each model of gravity.

To test this conjecture, Wald  devised a thought experiment, where  the charged, spinning test particles are dropped into the  extremal Kerr-Newman  BH \cite{ref3}. It was shown that the test particle could not be captured by the extremal Kerr-Newman  BH, implying the singularity can be hidden,  which is consistent  with the   WCCC. Later on, the relevant research was extended  to the scalar and electromagnetic test fields\cite{Semiz:2005gs,Duztas:2013wua,Duztas:2013gza,ref7}. Recently, the   WCCC   was   actively studied in many  BHs  via various methods\cite{ref8,ref9,ref10,ref11,ref12,ref13,ref14,ref15,ref16,ref17,ref18,ref19,ref20,ref21,ref22,ref23,ref24,ref25,ref26,ref27,ref28,ref29,add1,add2,add3,add4}.  Especially, Gwak investigated  the thermodynamics and   WCCC   of a charged AdS  Reissner-Nordstr$\ddot{o}$m    BH  with pressure and volume\cite{Gwak:2017kkt}.  He found that the SLT would be violated as the pressure was considered although the first law and    WCCC    still hold, which  is different  from the normal phase space without the pressure and volume contributions. Thereafter, using the test particle model of Gwak, thermodynamics and   WCCC   of a series   BHs have been investigated in the  EPS  \cite{Han:2019lfs,Wang:2019dzl,Han:2019kjr,Zeng:2019jrh,Zeng:2019jta,Zeng:2019aao}. Importantly, their results shown that the   WCCC    is valid for  the extremal  BHs in the  EPS  since the configurations of the  BHs are not changed as the particle is adsorbed. However, the   WCCC   of the near-extremal   BHs  is violable in the  EPS  \cite{He:2019kws,Zeng:2019hux,He:2019fti,Zeng:2019baw}.

In this paper, we shall investigate the thermodynamics and   WCCC   in generalized Rastall theories of gravity by making use of  the test fermion model in the  EPS . We want to explore whether the thermodynamics and   WCCC   will be valid in Rastall theories of gravity. The Rastall theory contains richer physics in comparison with the Einstein theory.
In General relativity, the geometry and matter fields are coupled minimally which results in the covariant conservation relation of the energy-momentum tensor. However, this relationship is only verified in the Minkowaski flat or weak field regime of gravity. Based on this limitation, the revised theory  of Einstein¡¯s gravity theory  was proposed by Rastall, in which the covariant derivative of the energy  momentum tensor is no longer zero in curved spacetime.   The thermodynamic behavior of  BHs in the Rastall gravity are more general, and can better reflect the nature of gravity as a thermodynamic system. Therefore, we want to explore the thermodynamic laws and  WCCC  in this  context.
As a result, the   FLT   is recovered by the absorptions of the fermion. However,  the entropy of   BH   decrease when a fermion drops into the extremal and near-extremal  BH, which violates the  SLT  . Fortunately, the result shows that the event horizon still exists when a fermion is swallowed by the extremal  BH. That is,  the extremal  BH in generalized Rastall theories of gravity can not be overcharged, and the   WCCC   is valid.
Different from the case of extremal, the   WCCC   for the near-extremal  BH could be invalid, depending on the model parameters of the Rastall theories of gravity.

The outline of the paper is organized as following. In section 2, we  will  review the   BH  solution and the thermodynamic quantities in the generalized Rastall theories of gravity. In section 3, we get the energy-momentum  relation of a charged fermion as it is swallowed by the  BH.  In section 4, the first and second law of  thermodynamics  are checked in the  EPS. In section 5, we concentrate on the   WCCC   under the fermion absorption with the contribution of the pressure. In section 6, we present our summarize.

\section{A brief review about charged AdS black hole in the Rastall gravity} \label{sec:2}
The covariant conservation of   energy-momentum tensor is  the backbone of the general relativity theory. Indeed, the limitation of such theory is that conservation of energy-momentum tensor has been probed only in the flat or weak field cases of spacetime.  Theoretically, in the strong domain of gravity,  the actual nature of the spacetime geometry and the covariant conservation relation should be  debated. In light of this fact,  by relating $T_{\text{   };\mu }^{\mu \nu }$ to the derivative of Ricci scalar,  Rastall proposed a new formulation for gravity by adding some new terms to the Einstein¡¯s equation\cite{ref41}. Therefore, on the basis of Rastall theory of gravity, the ordinary energy-momentum conservation law is not always available in the curved spacetime and  we should have\cite{Moradpour:2016fur}
\begin{equation}
T_{\text{     }\text{  };\mu }^{\mu \nu }={\lambda R}_{\text{     }}^{,\nu },\label{metric1}~~~~~~
\end{equation}
where $R$ is  the Ricci scalar of the spacetime, $\lambda$ is the Rastall  parameter. It leads to the modification of the Einstein¡¯s field equations, which can be written as
\begin{equation}
G_{\mu \nu }+\kappa  \lambda  g_{\mu \nu } R=\kappa  T_{\mu \nu }.\label{metric2}
\end{equation}
Here, $G_{\mu \nu }, T_{\mu \nu }$ and $\kappa$ are Einstein tensor, energy-momentum tensor and coupling constant respectively.  For case of  $\lambda = 0$, the Einstein field equations can be recovered. In recent years, the Rastall theory has attracted great attention and  many works on various   BH  solutions and  related thermodynamics have been investigated  in the framework of Rastall theory \cite{Moradpour:2016fur,Bronnikov:2016odv,Moradpour:2016rml,Heydarzade:2016zof,Heydarzade:2017wxu,Spallucci:2017mto,Lobo:2017dib,Kumar:2017qws,Lin:2018coh,Halder:2019akt }. For this theory, let us consider the field of matter consists only of electromagnetic field in the cosmological constant background,  and the static spherically symmetric metric would be  given as\cite{ref52}
\begin{equation}
{ds}^2=-h(r)d t^2+\frac{{dr}^2}{h(r)}+r^2 {d\Omega }_{d-2}^2,\label{metric3}
\end{equation}
and
\begin{equation}
h(r)=1-\frac{2M}{r}+\frac{Q^2}{r^2}-\frac{\Lambda }{3-12\lambda }r^2,\label{metric4}
\end{equation}
in which, $h(r)$ is the metric function, which is  determined in terms of  mass $M$ and charge $Q$, and ${d\Omega}_{d-2}^2={d\theta}^2+\sin^2{\theta d\phi }^2$  is  the volume of the unit $(d-2)$-sphere. The horizon radius is defined by $h(r) = 0$,  and the thermal properties  are defined on the horizon   of the  BH.  The Hawking temperature $T_h$,  entropy $S_h$, and electric potential $\Phi_h$ are obtained respectively
\begin{align}
&T_h=\frac{(1-4\lambda )Q^2+(4\lambda -1)r_h^2+\Lambda  r_h^4}{4(4\lambda -1)\pi  r_h^3},\label {Temperature}
\end{align}
\begin{align}
S_h=\pi  r_h{}^2,\label{Entropy}
\end{align}
\begin{align}
\Phi _h=-A_t\left(r_h\right)=\frac{Q}{r_h}.\label{Potential}
\end{align}
The cosmological constant $\Lambda$ is related to the AdS radius $l$, and both of them are related with the   pressure $P=-\frac{\Lambda }{8\pi }=\frac{(d-1)(d-2)}{16\pi  l^2}$. In the normal phase space, the cosmological constant is fixed. However, many researches have shown that the cosmological constant and its conjugate quantity can be identified as the thermodynamic pressure  and volume\cite{Caldarelli:1999xj,Dolan:2010ha,Cvetic:2010jb,Kastor:2009wy}. Therefore, the thermodynamic volume can be obtained by
\begin{equation}
V_h=\left(\frac{\partial M}{\partial P}\right)_{S, Q }=\frac{1}{1-4\lambda }\text{  }\left(\frac{4\pi  r_h{}^3}{3}\right).\label{Volume}
\end{equation}
The first law of the   BH  thermodynamics is now written by
\begin{align}
dM=T_{h}dS_{h}+\Phi_{h}dQ+V_{h}dP,  \label{firstlaw}
\end{align}
and corresponding Smarr relation  is also satisfied
\begin{equation}
M=2\left(T_hS_h-{PV}_h\right)+\Phi _hQ. \label{Smarr}
\end{equation}
When the cosmological constant is  treated as a constant, Eq.(\ref{firstlaw}) is reduced to  $dM= T_h dS_h+\Phi_{h}dQ$. In this paper, we are interested in the  EPS, that is, the Eq.(\ref{firstlaw}) is our concern. In this case, the mass $M$ is not the internal energy but enthalpy, which relates to the internal energy and  pressure, it is
\begin{align}
M=U_h+PV_h,  \label{eq2.15}
\end{align}
where $U_h$ is the internal energy. Next, we will investigate whether the first law can be recovered under the absorption of fermion.
\section{Charged fermion absorption}\label{sec3}
When the   BH  absorbs the charged fermion, the conserved quantities of the   BH  will change. In this process, the motion of a charged fermion following  the Dirac equation\cite{ref27}
\begin{align}
~~~~~~ \mathit{i}\gamma ^{\mu }\left(\partial _{\mu }+\Omega _{\mu }-\frac{\mathit{i}}{\hbar }e A_{\mu }\right)\Psi +\frac{\mu_m}{\hbar }\Psi =0,  \label{DiracEq}~~~~~~~~~
\end{align}
and
\begin{align}
~~~~~~\Omega _{\mu }\equiv \frac{\mathit{i}}{2}\omega _{\mu }{}^{\alpha \beta }\Sigma _{\alpha \beta },  \label{DiracEq1}~~~~~~
\end{align}
\begin{align}
~~~~~~\Sigma _{\alpha \beta }=\frac{\mathit{i}}{4}\left[\gamma ^{\alpha },\gamma ^{\beta }\right], \left\{\gamma ^{\alpha },\gamma ^{\beta }\right\}=2\eta ^{\alpha \beta }, \label{DiracEq2}~~~~~~
\end{align}
where ${\mu_m}$ and $e$ are the mass and charge of the fermion respectively. For the matrices $\gamma_\mu$, there are many different choice, and we can set
~~~~~~\begin{align}
&\gamma ^t=\frac{1}{\sqrt{h(r)}}\left(
\begin{array}{cc}
 \mathit{i} & 0 \\
 0 & -\mathit{i}~~~~~~
\end{array}
\right), \quad \gamma ^{\theta }=r\left(
\begin{array}{cc}
 0 & \sigma ^1 \\
 \sigma ^1 & 0
\end{array}
\right), \nonumber\\
&\gamma ^r=\frac{1}{\sqrt{h(r)}}\left(
\begin{array}{cc}
 0 & \sigma ^3 \\
 \sigma ^3 & 0
\end{array}
\right),\quad \gamma ^{\varphi }=r {\sin\theta} \left(
\begin{array}{cc}
 0 & \sigma ^2 \\
 \sigma ^2 & 0
\end{array}
\right). \label{Matrices}
\end{align}~~~~~~
In the above equation, we can employ the Pauli sigma matrices, which are
~~~~~~\begin{align}
\sigma ^1=\left(
\begin{array}{cc}
 0 & 1 \\
 1 & 0
\end{array}
\right),\quad  \sigma ^2=\left(
\begin{array}{cc}
 0 & -\mathit{i} \\~~~~~~
 \mathit{i} & 0
\end{array}
\right), \quad \sigma ^3=\left(
\begin{array}{cc}
 1 & 0 \\
 0 & -1
\end{array}
\right). \label{Pauli}~~~~~~
\end{align}~~~~~~
For the fermion with spin 1/2, there are two different states:  spin up ($\uparrow$) and spin down ($\downarrow$). Due to the spin down case  is similar to the spin up case. Here, we choose the wave functions with spin up ($\uparrow$) thereafter, this leading to
\begin{align}
\Psi _{(\uparrow )}=\left(
\begin{array}{c}
 A\\
 0\\
 B\\
 0\\
\end{array}~~~~~~
\right)e^{\left(\frac{\mathit{i}}{\hbar }I_{(\uparrow )}(t,r,\theta ,\phi )\right)},  \label{WaveEq}
\end{align}
in which, $A, B$, $I$ are functions of  $t, r, \theta, \phi$.
By combining Eqs. (\ref{WaveEq}) and  (\ref{DiracEq}),  we can obtain the motion equations to leading order of $\hbar$
\begin{align}
&-\mathit{i} \frac{A}{\sqrt{h(r)}}\left(\partial _tI_{(\uparrow )}-\text{eA}_t\right)-B\sqrt{h(r)}\partial _rI_{(\uparrow )}+A{\mu_m}=0, \nonumber\\
&-\mathit{i} \frac{B}{\sqrt{h(r)}}\left(\partial _tI_{(\uparrow )}-\text{eA}_t\right)-A\sqrt{h(r)}\partial _rI_{(\uparrow )}+B{\mu_m}=0, \nonumber\\
&~~~~~~~~~~~~~~~~~~A\left(r\partial _{\theta }I_{(\uparrow )}+\mathit{i} r {\sin\theta } \partial _{\phi }I_{(\uparrow )}\right)=0, \nonumber\\
&~~~~~~~~~~~~~~~~~~B\left(r\partial _{\theta }I_{(\uparrow )}+\mathit{i} r {\sin\theta } \partial _{\phi }I_{(\uparrow )}\right)=0
\label{WaveEq1}~~~~~~
\end{align}
We will pay more attention to the  first and second expressions in Eq.(\ref{WaveEq1}), since the radial action is determined by them. To solve these equations, we should separate the action $I_{(\uparrow )}$, it can be expressed as
\begin{align}
I_{(\uparrow )}=-{\omega t}+R(r)+\Theta (\theta ,\phi ).  \label{ActionI}
\end{align}
where $\omega$ is the energy of fermion which is near the  horizon. Putting Eq.(\ref{ActionI}) into Eq.(\ref{WaveEq1}) and canceling $A$ and $B$,
we can obtain the radial momentum which we are most interested in, that is
\begin{align}
p^r\equiv g^{{rr}}p_r=g^{{rr}}\partial _rI_{(\uparrow )}=\sqrt{\left(\omega +{eA}_t\right){}^2+{\mu_m}^2h(r)}.\label{radial}
\end{align}
The  charged fermion will be absorbed by the   BH  when it drops into the   BH  horizon.  In addition, near the event horizon, We have $h(r_h)\rightarrow0$ , which means
\begin{align}
\omega=\Phi_h e+ p^r_h,  \label{Radial}
\end{align}
in which $p^r_h$ is the radial momentum of the ingoing fermion near the  event horizon of   BH. Equation (\ref{Radial}) is a relation between the momentum, energy and charge of the ingoing fermion. when $\omega<\Phi_h e$,  the energy of the   BH  flows out the horizon,  which leads to the supperradiation occurs\cite{ref27}.  However,  when the charged fermion comes into the   BH  in the positive flow of time, the  energy  of the charged fermion should be defined as a positive value. Therefore, we will choose the positive sign in front of $p^r_h$ thereafter as done in Ref.\cite{ref57} in order to assure a positive time direction.

\section{Thermodynamic  with pressure and volume of the  BH under charged fermion absorption } \label{sec:4}
In this section, we will use the  relation (\ref{Radial}) to study the thermodynamics of  the  BH. Upon a charged fermion is absorbed by the  BH, the  change of    BH  is  infinitesimally  due to energy and momenta of the fermion entering into the event horizon of the  BH, so that the final state of the  BH is represented by ($M+dM, Q+dQ, r_h+dr_h, l+dl$), and $dM, dQ, dr_h, dl$ denote the increases of the mass, charge and radius respectively. In the process of absorption,  the corresponding  BH mass and electric charge infinitesimally change by the ingoing fermion, which is
\begin{align}
\omega=dU_h=d(M-PV_h), \quad e=dQ.  \label{InternalEg}
\end{align}
The relation (\ref{InternalEg}) is different the  form in the normal phase space where the internal energy is related to the mass of the  BH.
Then, the energy momentum relation in Eq.(\ref{Radial}) becomes as
\begin{align}
d(M-PV_h)=\Phi_h dQ+p^r_h. \label{EnergyEq}
\end{align}
To rewrite Eq.(\ref{EnergyEq}) to the   FLT  , we first use  Eq.(\ref{Entropy}), yielding the change of entropy
\begin{align}
{dS}_h=2 \pi r_h {dr}_h.  \label{DEntropy}
\end{align}
The variation of the event horizon $dr_h$ determined by the charge, energy, and radial momentum of the  absorbed fermion will directly contribute to the changes of $h(r+r_h)$. That is, the change of  $h(r)$ could lead to the shift of event horizon $({r_h}+{dr}_h)$.  Even so, the change of the horizon should satisfy
\begin{align}
d{h_r}=\frac{\partial h_r}{\partial M}{dM}+\frac{\partial h_r}{\partial Q}{dQ}+\frac{\partial h_r}{\partial r_h}{dr}_h=0,\quad h_r=h\left(M, Q, l, r_h\right),  \label{Horizon}
\end{align}
where
\begin{align}
&\frac{\partial h_r}{\partial M}|_{r=r_h}=-\frac{2}{r_h},   \quad \nonumber\\
&\frac{\partial h_r}{\partial Q}|_{r=r_h}=\frac{2 Q}{r_h{}^2},  \quad \nonumber\\
&\frac{\partial h_r}{\partial l}|_{r=r_h}=-\frac{2 r_h{}^2}{l^3 (1-4 \lambda )}, \quad \nonumber\\
&\frac{\partial h_r}{\partial r}|_{r=r_h}=-\frac{2 Q^2}{r_h{}^3}+\frac{2 M}{r_h{}^2}+\frac{2 r_h}{l^2 (1-4 \lambda )}.  \label{Horizon1}
\end{align}
With the help of Eq.(\ref{EnergyEq}) and Eq.(\ref{Horizon}), we  can get
\begin{align}
{dr}_h=\frac{2 l^2{  }r_h{}^2 (-1+4 \lambda )p_h^r}{r_h{}^4-2 l^2 \left(Q^2-M r_h\right) (-1+4 \lambda )}. \label{DHorizon}
\end{align}
Subsequently, with  the energy relation, the variations of entropy, and thermodynamic volume of the   BH   can be written  as
\begin{align}
{dS}_h=\frac{4 l^2 \pi  r_h{}^3 (-1+4 \lambda )p_h^r}{r_h{}^4-2 l^2 \left(Q^2-M r\right) (-1+4 \lambda )},  \label{DEntropy1}
\end{align}
\begin{align}
{dV}_h=-\frac{8 l^2 \pi \text{  }r_h{}^4p_h^r}{r_h{}^4-2 l^2 \left(Q^2-M r_h\right) (-1+4 \lambda )}.  \label{DVolume}
\end{align}
Making use of Eqs.(\ref{Temperature}), (\ref{DEntropy1}) and Eq.(\ref{DVolume}),  we find a relation, which is
\begin{align}
T_h{dS}_h-{PdV}_h=p_h^r.  \label{eq4.10}
\end{align}
Thus, the expression of the internal energy in Eq.(\ref{EnergyEq})  becomes
\begin{align}
dM=\Phi_h dQ + T_h dS_h+V_h dP.  \label{eq4.11}
\end{align}
Therefore, it is worth noting that the coincidence between the variation of the  charged   BH  in the Rastall gravity and the   FLT   under the fermion absorption, where  the cosmological constant is treated as a dynamical variable.

With  Eq.(\ref{DEntropy1}), we  can investigate  the   SLT   in the generalized Rastall theories. For the extremal  BHs where $T_h=0$, we can get
\begin{equation}
{dS}_{extremal}=\frac{4 l^2 \pi {  }(-1+4 \lambda )p_h^r}{3 r_h}. \label{EntropyEx}
\end{equation}
In Ref.\cite{ref52}, for  case of $\Lambda<0$, $\lambda<1/4$ corresponds to the Anti-de Sitter spacetime, so we only talk about $\lambda < 1/4$ in the following discussion.  Hence, there is  $-1+4 \lambda<0$ in the Eq.(\ref{EntropyEx}), which means we can get ${dS}_{extremal}<0$. That is to say, the  SLT   can be violated  at least for the extremal case  under the charged fermion absorption.

Next, we turn to the non-extremal  BH.  To gain an intuitive understanding, we plot $dS_h$ for different $\lambda$   and $M$  in Figure 1.
From this figure,  we find  there  is always a  divergent point  which divides $dS_h$ into positive and negative regions when we chose different values of $M$ and $\lambda$. Obviously, the negative region  means that the change of entropy is less than zero $(dS_h<0)$, and the decrease of the entropy appears in ranges close to the extremal  BH. Therefore,  the  SLT   is invalid under charged fermion absorption, and the range of the violation appears in extremal and  near-extremal condition. In addition, we also find that the violations of the second law depending on the model parameters $M, l, Q, \lambda$,  that is, the magnitudes of the violations are related to these parameters.
\begin{figure}[h]
 {\resizebox{0.45\textwidth}{!}{\includegraphics{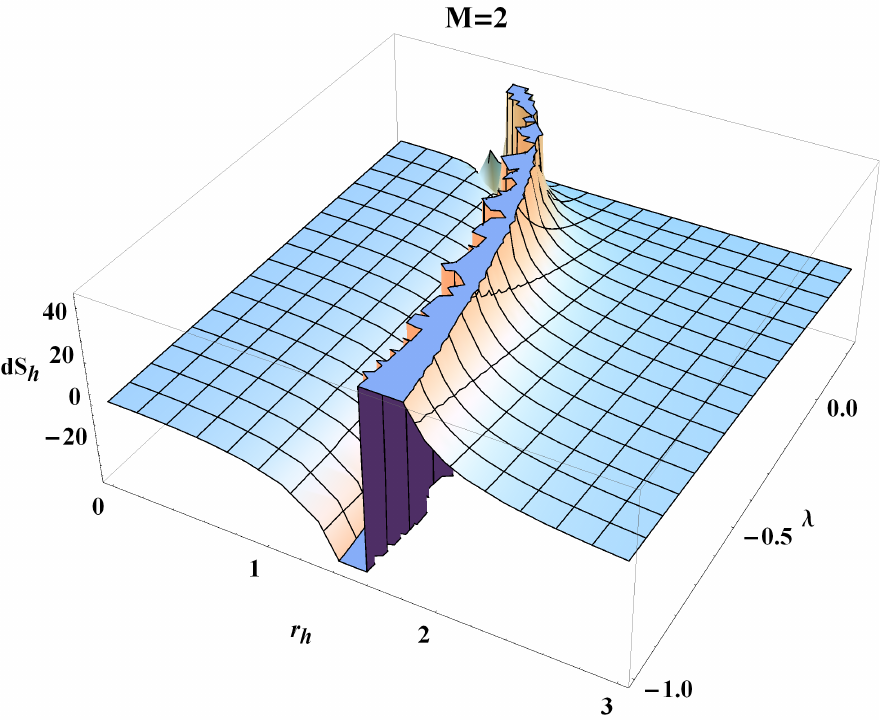}}}
 {\resizebox{0.45\textwidth}{!}{\includegraphics{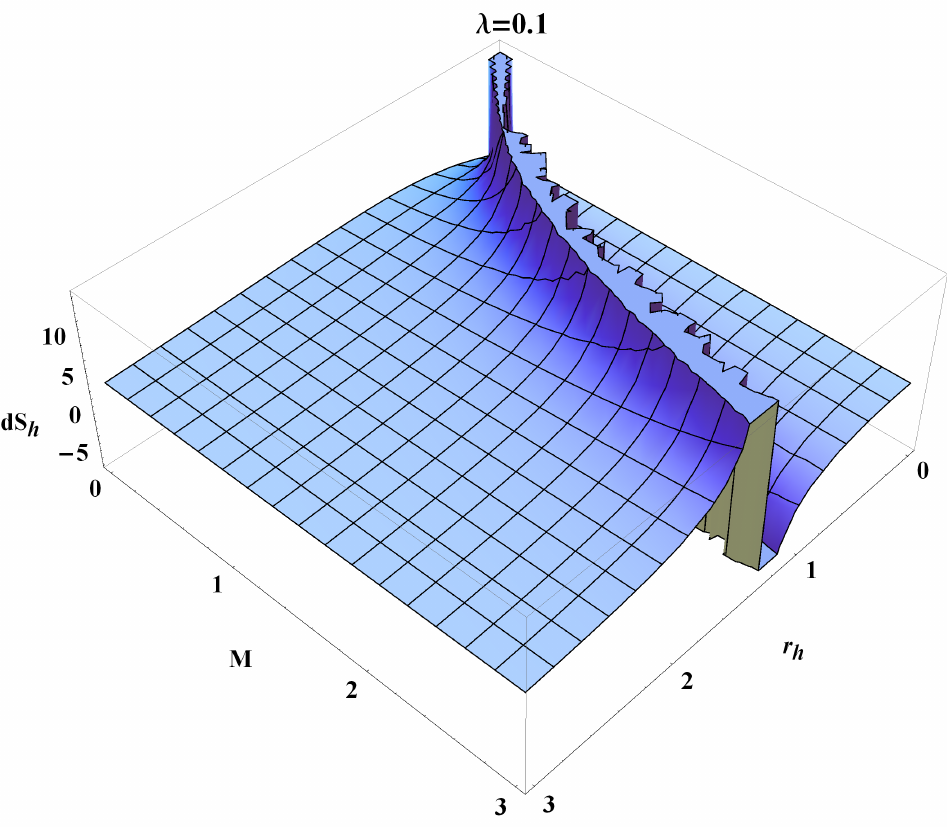}}}
 \caption{The figures of critical $dS_h$ for different $\lambda$ (left) and $M$ (right) when $p_h^r=1$.}
\label{fig:3}
\end{figure}

\section{Weak cosmic censorship conjecture with pressure and volume }
\label{sec:5}
When the absorbed charge is  enough more, the   BH  is overcharged and the   WCCC   is violated. In this section, we will study the validity of the   WCCC   in the process that  the fermions are swallowed  by  BHs. Therefore, we need check whether the  event horizon will be  destroyed at the final state. If the event horizon exist, the metric  function $h(r)$ of the    BH  has $h(r_{\min })\leq0$ where $r_{\min}$ is a  minimum point of the function.  At $r_{\min }$, the following relations are satisfied
\begin{align}
h(r)|_{r=r_\text{min}}\equiv h_\text{min}=\delta\leq 0,\quad \partial_{r}h(r)|_{r=r_\text{min}}\equiv h'_\text{min}=0, \quad
(\partial_{r})^2 h(r)|_{r=r_\text{min}}>0.  \label{Minimum0}
\end{align}
When the minimum value satisfies the condition  $\delta\doteq 0$,  the   BH  is  an extremal  BH. There, $h(r)$ is a function of ($M, Q, l$), and it becomes to $h(M+dM, Q+dQ, l+dl)$ during  the process that a fermion drops into the event horizon.  Correspondingly, the position of the minimum point and event horizon should change into $r_{\min}\rightarrow r_{\min}+dr_{\min}, r_h\rightarrow r_h+dr_h$. After then, $h(r)$ also have a small shift which is denoted  by $dh_{\min}$. At the new lowest point , we have
\begin{align}
\partial _r h|_{r=r_{\min }+{dr}_{\min }}=h'_{\min }+{dh'}_{\min }=0.  \label{lowestpoint}~~~
\end{align}
\begin{align}
{dh}'_\text{min }=\frac{\partial h'_\text{min }}{\partial M} dM+\frac{\partial h'_\text{min }}{\partial Q} dQ+\frac{\partial h'_\text{min }}{\partial r_\text{min }} dr_\text{min }+\frac{\partial h'_\text{min }}{\partial l} dl=0.  \label{Minimum}
\end{align}
Then, at the new minimum point, we can get
\begin{align}
h\left(r_\text{min }+dr_\text{min }\right)=h_\text{min }+dh_\text{min }, \label{Minimum1}
\end{align}
where
\begin{align}
dh_\text{min }=\frac{\partial h_\text{min }}{\partial M} dM+\frac{\partial h_\text{min }}{\partial Q} dQ +\frac{\partial h_\text{min }}{\partial l} dl. \label{Minimum2}
\end{align}
Inserting the condition (\ref{Minimum0}) into Eq. (\ref{Minimum2}),  and combining with Eq.(\ref{EnergyEq}) we obtain
\begin{align}
dh_\text{min }=0. \label{DMinimum}
\end{align}
For the extremal   BH  where  we have $h_\text{min }=\delta =0$,   the transformation of $h\left(r_\text{min }+dr_\text{min }\right)$ is expressed as
\begin{align}
h_\text{min }+dh_\text{min }=0.  \label{Value1}
\end{align}
This means that the event  horizon always exists, which depicts that the   WCCC   is valid  in the  EPS .
It is interesting to note that the extremal   BH  keeps its configuration after the absorption. In other words,  the extremal   BH  is still extremal, that is,  the fermion with sufficient momentum and charge would not overcharge extremal  BH in the  Rastall  gravity.

For the near-extremal  BH, the energy of the fermion in Eq.(\ref{EnergyEq}) can not be used because it is just applicable at the event horizon. Hence, we can expand Eq. (\ref{EnergyEq}) at $r_\text{min }+\epsilon$, which is
\begin{align}
&{dM}=\frac{2 r _\text{min }\left( r_\text{min }{}^4{dl}+ l^3 Q (4 \lambda -1){dQ}\right)-\left(3 l r_\text{min }{}^4+l^3 \left(Q^2-r_\text{min }{}^2\right) (4 \lambda -1)\right){dr}_\text{min }}{2 l^3 r_\text{min }{}^2 (4 \lambda -1)}  \quad \nonumber\\
&~~~~+\frac{\left(3 r_\text{min }{}^5{dl}+l^3 Q r_\text{min } (1-4 \lambda ){dQ} +\left(l^3 Q^2 (4 \lambda -1)-3 l r_\text{min }{}^4\right){dr}_\text{min }\right) \epsilon }{l^3 r_\text{min }{}^3 (4 \lambda -1)}+O(\epsilon)^2. \label{DMass}
\end{align}
By combining Eqs.(\ref{DMass}) and (\ref{Minimum2}) we have
\begin{align}
&{dh}_{\min }=\frac{ \left(3 r_\text{min }{}^4+l^2 \left(Q^2-r_\text{min }{}^2\right) (4 \lambda -1)\right){dr}_\text{min }}{l^2 r_\text{min }{}^3 (4 \lambda -1)} \quad \nonumber\\
&~~~~~~~~-\frac{2 \left(3 r_\text{min }{}^5{dl}+l^3 Q r_\text{min } (1-4 \lambda ){dQ}+\left(-3 l r_\text{min }^4+l^3 Q^2 (-1+4 \lambda )\right){dr}_\text{min }\right) \epsilon }{l^3 r_\text{min }{}^4 (4 \lambda -1)}+O(\epsilon)^2. \label{DMinimumN}
\end{align}
Meanwhile, we get the expression of $l$ with the help of $h'(r_h)=0$, which is
\begin{equation}
l=\frac{\sqrt{3} r_\text{min }{}^2}{\sqrt{\left(-Q^2+r_\text{min }{}^2\right) (4 \lambda -1)}}, \label{LValue}
\end{equation}
and
\begin{align}
{dl}=\frac{\sqrt{3} r_\text{min } \left( Q r_\text{min } {dQ}+{dr}_\text{min } \left(-2 Q^2+r_\text{min }{}^2\right)\right) (4 \lambda -1)}{\left(\left(-Q^2+r_\text{min }{}^2\right) (4 \lambda -1)\right){}^{3/2}}. \label{DLValue}
\end{align}
Substituting (\ref{DMinimumN}), (\ref{LValue}) and (\ref{DLValue}), we can get
\begin{align}
{dh}_{\min }=O(\epsilon )^2.   \label{NMM}
\end{align}
In the  EPS, the minimum value of the near-extremal   BH  is
\begin{align}
h_{\min }+{dh}_{\min }=\delta _{\epsilon }+O(\epsilon )^2.  \label{NMM1}
\end{align}
Naturally, once again we can obtain $h_{\min}+dh_{\min}=0$ when $\delta_{\epsilon}\rightarrow0$, $\epsilon\rightarrow0$, such that the accuracy of Eq.(\ref{Value1}) is verified. However,  both $\delta _{\epsilon }$ and $O(\epsilon )^2$ are small quantities about $\epsilon$, and we can not directly judge the size of these small quantities. Therefore, we want find the  specific expression of  $O(\epsilon )^2$ and $\delta _{\epsilon }$ to discuss which is smaller.
For  the near-extremal  BH, we  should  perform higher-order expansion, namely
\begin{align}
{dh}_M=\frac{\left(6 Q^2 {dr}_\text{min }-4 Q r_\text{min }{dQ}- r_\text{min }{}^2 {dr}_\text{min }\right) \epsilon ^2}{r_\text{min }{}^5}+O(\epsilon )^3. \label{NMM3}
\end{align}
For $\delta _{\epsilon }$, we also have
\begin{align}
\delta _{\epsilon }=\frac{\left(-3 r_\text{min }{}^4+l^2 Q^2 (4 \lambda -1)\right)\epsilon ^2}{l^2 r_\text{min }{}^4 (4 \lambda -1)}+O(\epsilon )^3.\label{Tda3}
\end{align}
Thus, we can combine  equation (\ref{NMM3}) and equation (\ref{Tda3}) and define
\begin{align}
\mathcal{H}_E=\frac{\delta _{\epsilon }+{dh}_M}{\epsilon ^2}. \label{Finally}
\end{align}
Here, the next step is to analyze the sign of Eq.(\ref{Finally}). If the   WCCC   of the near-extermal   BH  is valid in the  EPS , the value of Eq.(\ref{Finally}) should be negative $(\mathcal{H}_E<0)$. Apparently, the value of $\mathcal{H}_E$ depends on the model parameters $( r_{\min}, Q, l, \lambda, dr_{\min})$. As an attempt, we set $Q = 2, l = 1, dQ = 0.5, p_h^r=1$, and plot the figure of Eq.(\ref{Finally}),  which makes the result  more intuitive.
\begin{figure}[h]
 {\resizebox{0.45\textwidth}{!}{\includegraphics{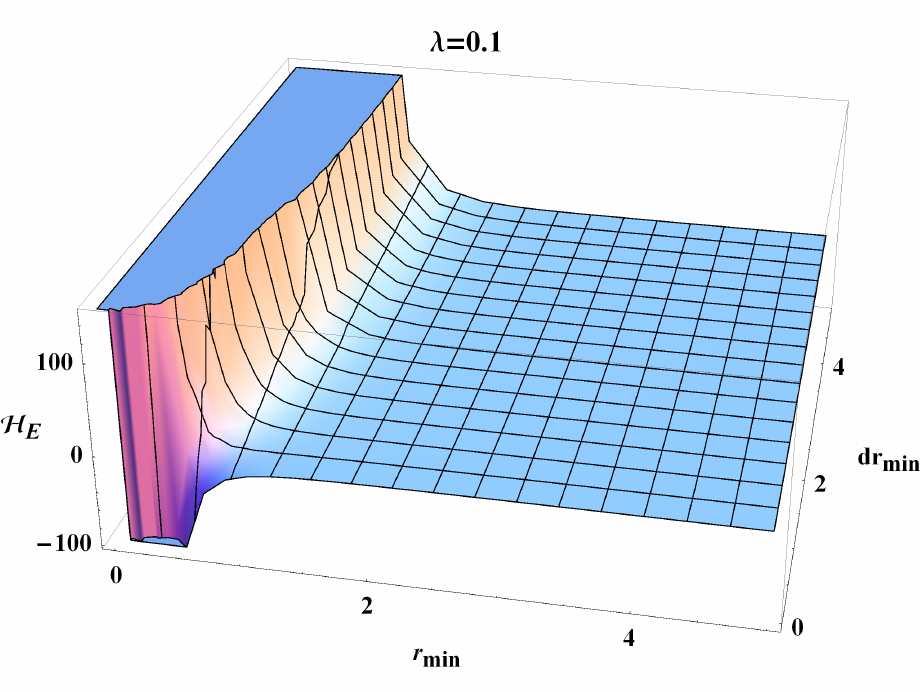}}}
 {\resizebox{0.45\textwidth}{!}{\includegraphics{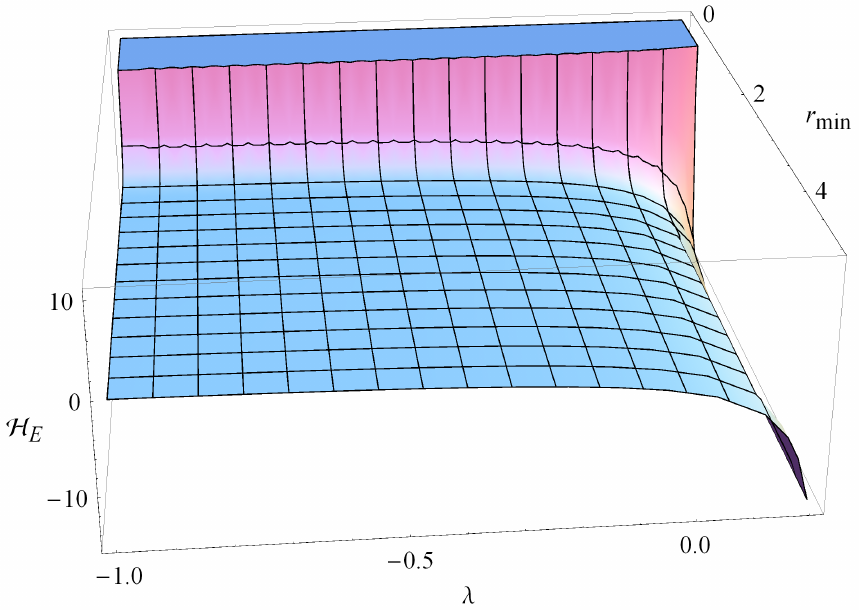}}}
 \caption{The value of $\mathcal{H}_E$  for different  $dr_\text{min }$ (left) and $\lambda$ (right) when $p_h^r=1$.}
\label{fig:3}
\end{figure}

In Figure 2, we take $\lambda = 0.1$ (left) and $dr_{\min} =0.5$ (right) respectively, it is worthwhile to point that there is a region where  $\mathcal{H}_E>0$ when we chose different values of ${dr}_{\min}$ or $\lambda$. Since the value of  $\mathcal{H}_E$ could be positive $(\mathcal{H}_E>0)$, which states that  the function $\mathcal{H}_E=h\left(r_\text{min }+dr_\text{min }\right)$ has no real root. In this case,  the   WCCC   could be not valid for the near-extermal  BH, and the near-extremal   BH  could be overcharged in the process. It is also obvious from the Figure 2,  the magnitudes of the violation  is related to  the parameters of the  BH, that is,  the magnitude  of the violation depends on the the  parameters $\lambda, {dr}_{\min}, r_{\min}$, which is  quite different from  the case of extremal  BH.

\section{Discussion and conclusions}\label{sec:5}
Thermodynamic laws and WCCC for a   BH  in Rastall gravity are discussed in the  EPS  where  the cosmological constant was treated as a dynamical variable.
We first obtained  the energy-momentum relation of the  particle which is  captured by   BHs in the Rastall gravity. Using  this relation, we  further studied the thermodynamic laws and  WCCC  of  BH. Fortunately, the   FLT   was found to be valid in the  EPS  under the fermion absorption. However, the change in the entropy of extremal   BH  was found to be  less than zero $({dS}_{extremal}<0)$, implying the  SLT   of extremal   BH  is invalid in the extended  phase space. For  the non-extremal  BH, we found that the  SLT   may be violated  for near-extremal  BHs, and the magnitudes of the violation rely on the choice of parameter model. This  conclusion is consistent with the results in Einstein gravity \cite{Gwak:2017kkt}.

It is well known that the  Bekenstein-Hawking entropy of the BH is related to the event horizon, that is,  the thermodynamics of a  BH is related to the stability of its horizon. Since there is a violation of the  SLT   of  BH, it is necessary to judge the existence of the horizon and further check the WCCC. To determine whether the event horizon still exists when a fermion falls in the extremal  BH,  the most effective method   is to check the minimum value of function $h(r)$. Interestingly,  we found that the minimum value of the function was not changed  for the extremal  BH, that is  $(dh_\text{min }=0)$, so that  $h_\text{min }+dh_\text{min }=0$,  implying that the extremal  BH cannot be overcharged in the process due to the its minimum value. The extremal  BH is still extreme. This ensures the stability of the horizon under the charged particle absorption. Therefore, the conjecture showed its validity for the extremal    BH  in the Rastall gravity by adding a fermion. However, there is a situation where the minimum value of the function is greater than zero for the near-extremal  BH $(\mathcal{H}_E>0)$, which was shown in Figure 2. Clearly, for case of the near-extremal  BH, the   WCCC   could be violable, similarly, the magnitudes of the violation depend on the values of related parameters.

\vspace{10pt}

\noindent {\bf Acknowledgments}

\noindent
This work is supported  by the National
Natural Science Foundation of China (Grant Nos. 11875095, 11903025), and Basic Research Project of Science and Technology Committee of Chongqing (Grant No. cstc2018jcyjA2480).

\end{document}